\documentclass[12pt]{article}
\usepackage{bbm,latexsym}
\usepackage{amssymb,amsmath}
\usepackage{epsfig}
\newcommand{\be}{\begin{equation}}
\newcommand{\ee}{\end{equation}}
\newcommand{\ba}{\begin{eqnarray}}
\newcommand{\ea}{\end{eqnarray}}
\newcommand{\bs}{\begin{subequations}}
\newcommand{\es}{\end{subequations}}
\newcommand{\no}{\nonumber\\}

\newcommand{\diag}{\mbox{diag}}
\newcommand{\bone}{\mathbbm{1}}
\newcommand{\zz}{\mathbbm{Z}}
\newtheorem{theorem}{Theorem}
\begin{document}
\title{
\normalsize \hfill UWThPh-2017-27
\\[.5mm]
\normalsize \hfill CFTP/17-006
\\[4mm]
\LARGE Cobimaximal lepton mixing from \\ soft symmetry breaking}

\author{
W.~Grimus$^{(1)}$\thanks{E-mail: \tt walter.grimus@univie.ac.at}
\ and
\addtocounter{footnote}{2}
L.~Lavoura$^{(2)}$\thanks{E-mail: \tt balio@cftp.tecnico.ulisboa.pt}
\\*[3mm]
$^{(1)} \! $
\small University of Vienna, Faculty of Physics,
\small Boltzmanngasse 5, A--1090 Vienna, Austria
\\[2mm]
$^{(2)} \! $
\small Universidade de Lisboa, Instituto Superior T\'ecnico, CFTP,
\small 1049-001 Lisboa, Portugal
\\*[2mm]
}

\date{August 31, 2017}

\maketitle

\begin{abstract}
Cobimaximal lepton mixing, \textit{i.e.}\ 
$\theta_{23} = 45^\circ$ and $\delta = \pm 90^\circ$
in the lepton mixing matrix $V$,
arises as a consequence of $S V = V^\ast \mathcal{P}$,
where $S$ is the permutation matrix that interchanges
the second and third rows of $V$ and $\mathcal{P}$
is a diagonal matrix of phase factors.
We prove that any such $V$ may be written in the form $V = U R P$,
where $U$ is \emph{any predefined}\/ unitary matrix satisfying $S U = U^\ast$,
$R$ is an orthogonal,
\textit{i.e.}\ real,
matrix,
and $P$ is a diagonal matrix satisfying $P^2 = \mathcal{P}$.
Using this theorem,
we demonstrate the \emph{equivalence}\/ of two ways of constructing
models for cobimaximal mixing---one way that uses a standard $CP$ 
symmetry and a different way that uses a $CP$ symmetry
including $\mu$--$\tau$ interchange.
We also present two simple seesaw models to illustrate this equivalence;
those models have,
in addition to the $CP$ symmetry,
flavour symmetries broken softly
by the Majorana mass terms of the right-handed neutrino singlets.
Since each of the two models needs four scalar doublets,
we investigate how to accommodate the Standard Model
Higgs particle in them.
\end{abstract}

\newpage

\section{Introduction}

Cobimaximal lepton
mixing~\cite{fukuura,harrison,GL2003,joshipura,ma2016,ma2017},
\textit{i.e.}\ $\theta_{23} = 45^\circ$ and $\delta = \pm 90^\circ$,
where $\theta_{23}$ is the atmospheric mixing angle
and $\delta$ is the Dirac phase
in the lepton mixing (PMNS) matrix,
is still a viable option---for recent fits to the neutrino oscillation data
see ref.~\cite{esteban}.
Although the best-fit value value of $\theta_{23}$ is slightly off $45^\circ$,
$\theta_{23} = 45^\circ$ is still possible at the $3 \sigma$ level.
Moreover,
the most recent data prefer $\delta = -90^\circ$ to $\delta = +90^\circ$.
Cobimaximal mixing is equivalent to the condition
\be \label{abs}
\left| V_{\mu j} \right| = \left| V_{\tau j} \right| \quad \forall \, j=1,2,3
\ee
in the lepton mixing matrix $V$~\cite{harrison}. 
The condition~\eqref{abs} does not restrict
the mixing angles $\theta_{12}$ and $\theta_{13}$.
In model-building practice,
condition~\eqref{abs} is more easily achieved through the
condition
\be \label{SV}
S V = V^\ast \mathcal{P},
\ee
where $S$ is the permutation matrix given by
\be \label{S}
S = \left(
\begin{array}{cccc} 1 & 0 & 0 \\ 0 & 0 & 1 \\ 0 & 1 & 0 
\end{array} \right)
\ee
and $\mathcal{P}$ is a diagonal matrix of phase factors.
Notice that $S$ is real and unitary and that it satisfies
$S^2 = \bone$. (In this paper, $\bone$ denotes the $3 \times 3$ unit matrix.)

In the model-building literature one finds
two approaches to cobimaximal mixing
which can be distinguished by the $CP$ transformation property
of the Majorana mass Lagrangian of the three light neutrinos.
The first approach~\cite{harrison,GL2003,joshipura}
is based on the invariance of that mass Lagrangian under
a non-standard $CP$ transformation including the $\mu$--$\tau$ interchange;
the second approach~\cite{fukuura,ma2016,ma2017}
uses the standard $CP$ transformation.
In section~\ref{cobi} of this paper we demonstrate
that the two approaches are completely equivalent 
in their consequences for lepton mixing.
In sections~\ref{simple model1} and~\ref{simple model2} we present
two very simple models for cobimaximal mixing
that illustrate the equivalence of both approaches.
Those models include several Higgs doublets;
it is thus non-trivial to accommodate in them a Standard Model (SM)-like
Higgs boson with mass $m_h = 125$\,GeV~\cite{rpp};
a discussion of this issue is found in section~\ref{accommodation}.
Section~\ref{concl} presents our conclusions.
In order to facilitate the reading of sections~\ref{simple model1}
and~\ref{simple model2},
we display the general formulas for weak-basis changes in appendix~\ref{bt}.
Appendix~\ref{VEVa} discusses,
in a general multi-Higgs-doublet model, 
the conditions for the alignment of the vacuum expectation values (VEVs)
needed to accommodate the SM Higgs boson.

\section{Two equivalent approaches to cobimaximal mixing}
\label{cobi}

In order to fix the notation we write down the mass terms
\be
\label{mass}
\mathcal{L}_\mathrm{mass} =
- \bar \ell_L M_\ell \ell_R
+ \frac{1}{2}\, \nu_L^T C^{-1} M_\nu \nu_L
+ \mbox{H.c.}
\ee
for the chiral charged-lepton and neutrino fields $\ell$ and $\nu$,
respectively.
In equation~(\ref{mass}),
$C$ is the charge-conjugation matrix in Dirac space.
The mass matrices in flavour space $M_\ell$ and $M_\nu$
are both $3 \times 3$ in order to accommodate the three families;
$M_\nu$ is a symmetric but in general complex matrix
since it corresponds to Majorana mass terms.
The diagonalization of the mass matrices proceeds via
\bs
\ba
U_{\ell L}^\dagger M_\ell U_{\ell R} &=& \diag \left( m_e, m_\mu, m_\tau \right) \equiv
\hat m_\ell,
\\
U_\nu^T M_\nu U_\nu &=& \diag \left( m_1, m_2, m_3 \right) \equiv \hat m_\nu.
\ea
\es
The PMNS matrix is $V = U_{\ell L}^\dagger U_\nu$.

We now prove
\begin{theorem} \label{V}
  Any two $3 \times 3$ unitary matrices $U_1$ and $U_2$
  that fulfil $U_k^\ast = S U_k$ ($k = 1, 2$)
  differ only by an orthogonal matrix $R$,
  \textit{i.e.}\ $U_2 = U_1 R$.
\end{theorem}
\textbf{Proof:}
$U_1^\dagger U_2 = U_1^\dagger \left( S^\dagger S \right) U_2
= \left( S U_1 \right)^\dagger \left( S U_2 \right)
= U_1^T U_2^\ast
= \left( U_1^\dagger U_2 \right)^\ast$.
Thus,
$U_1^\dagger U_2 \equiv R$ is a real matrix,
hence it is orthogonal. Q.E.D.

\vspace*{3mm}

With theorem~\ref{V} it is easy to understand that the following two
approaches to cobimaximal mixing found in the literature are equivalent.
\begin{enumerate}
\item In the first approach~\cite{harrison,GL2003,joshipura},
  the charged-lepton mass matrix is diagonal
  (or,
  more generally,
  of the form $M_\ell = \hat m_\ell U_{\ell R}^\dagger$
  with diagonal $\hat m_\ell$ and unitary $U_{\ell R}$)
  while the Majorana neutrino mass term enjoys
  the generalized $CP$ symmetry\footnote{For the sake of clarity,
    we spell out the transformation~\eqref{bugiou} at length:
  \[
  \nu_{eL} \left( x_0, \vec x \right) \to i \gamma_0 C
  \bar \nu_{eL}^T \left( x_0, -\vec x \right), \
  \nu_{\mu L} \left( x_0, \vec x \right) \to i \gamma_0 C
  \bar \nu_{\tau L}^T \left( x_0, -\vec x \right), \
  \nu_{\tau L} \left( x_0, \vec x \right) \to i \gamma_0 C
  \bar \nu_{\mu L}^T \left( x_0, -\vec x \right).
  \]
}
  \be \label{bugiou}
  \nu_L \left( x^0, \vec x \right) \to i S \gamma_0 C
  \bar \nu_L^T \left( x^0, -\vec x \right).
  \ee
  Because of the symmetry~\eqref{bugiou},
  the mass matrix $M_\nu$
  fulfils
  the condition
  \be \label{mu-tau}
  S M_\nu S = M_\nu^\ast.
  \ee
  It has been shown in ref.~\cite{GL2003} that equation~\eqref{mu-tau} entails
  $U_\nu = U P$,
  where $U$ has the form of theorem~\ref{V},
  \textit{i.e.}\ $S U = U^\ast$,
  and $P$ is a diagonal matrix of phase factors.
  (Moreover,
  if $m_j \neq 0$ then $P_{jj}$ can only be either $\pm 1$ or $\pm i$.)
  Now,
  since the charged-lepton mass matrix is diagonal,
  the lepton mixing matrix $V$ coincides---apart
  from multiplication from the left
  with a diagonal matrix of phase factors---with $U_\nu$ and 
  is thus given by
  \be \label{second}
  V = U P.
  \ee
  Equation~(\ref{abs}) is then
  fulfilled
  and cobimaximal mixing obtained.
  Note that equation~\eqref{second} together with $S U = U^\ast$ lead to
  $S V = V^\ast P^2$,
  which is a special case of equation~\eqref{SV}.
\item
  In the second approach~\cite{fukuura,ma2016,ma2017},
  $M_\nu$ is real,
  \textit{i.e.}\ the standard $CP$ symmetry
  \be \label{cp2}
  \nu_L \left( x^0, \vec x \right) \to i \gamma_0 C
  \bar \nu_L^T \left( x^0, -\vec x \right)
  \ee
  applies to the neutrino Majorana mass terms.
  Furthermore,
  there is some symmetry in the charged-lepton mass terms
  such that either $U_{\ell L}^\dagger = U_\omega$,
  where
  \be\label{Uomega}
  U_\omega = \frac{1}{\sqrt{3}} \left( \begin{array}{ccc}
    1 & 1 & 1 \\
    1 & \omega & \omega^2 \\
    1 & \omega^2 & \omega
  \end{array} \right)
  \quad \mbox{with} \
  \omega = \exp{\left( 2 i \pi / 3 \right)},
  \ee
  or $U_{\ell L}^\dagger = U_\varrho$,
  where~\cite{ma2017}\footnote{In ref.~\cite{ma2017}
    the notation $U_2$ is used for our $U_\varrho$.}
  \be\label{Urho}
  U_\varrho
  = \left( \begin{array}{ccc}
    1 & 0 & 0 \\
    0 & \varrho & - i \varrho \\
    0 & \varrho & i \varrho
  \end{array} \right)
  \quad \mbox{with} \
  \varrho = \frac{1}{\sqrt{2}}.
  \ee
  Then,
  in both cases the PMNS matrix has the form
  \be\label{first}
  V = U_a R_a P_a \quad (a=\omega,\varrho),
  \ee
  with $R_a$ orthogonal and $P_a$ having the same properties as $P$  
  in the previous paragraph.
  (The matrix $U_\nu = R_a P_a$ diagonalizes
  the real matrix $M_\nu$.\footnote{The matrices $P_a$
    are needed in order to obtain positive neutrino masses $m_{1,2,3}$.
    Indeed,
    if $M_\nu$ is real then it is diagonalized by the orthogonal matrix $R_a$,
    but the diagonal matrix resulting from that diagonalization
    may contain some negative diagonal entries;
    when $m_j < 0$ one needs $\left( P_a \right)_{jj} = \pm i$
    to correct for that.})
  Since $R_a$ is real
  and the second and third rows in both $U_\omega$ and $U_\varrho$ 
  are the complex conjugates of each other,
  equation~\eqref{abs} holds 
  and the matrix $V$ of equation~(\ref{first})
  displays cobimaximal mixing.
\end{enumerate}

Now,
theorem~\ref{V} tells us that
both approaches lead to the same predictions for the mixing matrix,
namely cobimaximal mixing and Majorana phases 
in neutrinoless $\beta\beta$-decay which are either zero or $180^\circ$,
while the mixing angles $\theta_{12}$ and $\theta_{13}$
are not restricted.\footnote{We stress that this statement only refers to
  the conditions layed out in this section. In specific models the mixing
  matrix could be more predictive.} 
Indeed,
consider the first approach,
leading to the mixing matrix of equation~\eqref{second}.
Then,
from $S U = U^\ast$, we deduce with theorem~\ref{V}
that there are orthogonal matrices $R$ and $R'$ such that 
$U = U_\omega R = U_\varrho R'$.
Thus, equation~\eqref{second} may
be converted into the form of equation~\eqref{first} and both approaches are,
therefore,
equivalent.

If both approaches refer to \emph{the same model}, then the mixing matrices
in equations~\eqref{second} and~\eqref{first} are the same 
because
they just correspond to
\emph{the same $CP$ symmetry}\/ in that model but written
in different weak bases.
For the general formulas of a weak-basis change,
of a $CP$ transformation,
and of the transformation property under a weak-basis change
of the $CP$-transformation matrix in flavour space,
we refer the reader to
equations~\eqref{bughihjp},
\eqref{buigfp},
and~\eqref{bufiy},
respectively, in appendix~\ref{bt}.
In particular,
if $S_D$ is the matrix in flavour space that operates
the $CP$ transformation of the left-handed lepton doublets, 
we see that the $CP$ transformation~\eqref{bugiou}
corresponds to $S_D = S$.
According to equation~\eqref{bufiy},
a change of weak basis performed by a unitary matrix $W_D$
satisfying $W_D^\ast = S W_D$
changes $S_D = S$ to
\be\label{bone}
S^\prime_D
= W_D^\dagger S_D W_D^\ast
= W_D^\dagger S W_D^\ast
= W_D^\dagger S \left( S W_D \right) = W_D^\dagger W_D = \bone,
\ee
\textit{i.e.}\ the $CP$ transformation becomes
the standard $CP$ transformation of equation~\eqref{cp2}.
Now,
since $U_\omega^\ast = S U_\omega$ and $U_\varrho^\ast = S U_\varrho$,
both $U_\omega$ and $U_\varrho$
are well suited to perform changes of weak basis that lead
from $S_D = S$ to $S^\prime_D = \bone$.
The resulting PMNS matrix $V = U P$ of equation~\eqref{second}
cannot change when the weak basis is changed,
hence,
since $S U = U^\ast$,
one must have $U = U_\omega R_\omega = U_\varrho R_\varrho$
as in equation~\eqref{first};
this is precisely what theorem~\ref{V} guarantees us.

\section{A simple model for cobimaximal mixing}
\label{simple model1}

We construct an extension of the SM with gauge group $SU(2) \times U(1)$
and \emph{four} scalar $SU(2)$ doublets
$\phi_\alpha$ ($\alpha = e, \mu, \tau$) and $\phi_\nu$.
The Yukawa Lagrangian is
\be
\label{LY}
\mathcal{L}_Y = - \sum_{\alpha=e,\mu,\tau} \bar D_{\alpha L} 
\left( y_1 \phi_\alpha \alpha_R + 
y_2 \tilde\phi_\nu \nu_{\alpha R} \right) + \mbox{H.c.}
\ee
In equation~\eqref{LY},
the $D_{\alpha L} = \left( \nu_{\alpha L},\ \alpha_L \right)^T$
are the leptonic gauge-$SU(2)$ doublets,
the $\alpha_R$ are the right-handed charged-lepton fields,
and the $\nu_{\alpha R}$ are three right-handed neutrinos
that we have introduced in order to enable a seesaw mechanism.
The Lagrangian~\eqref{LY} has only two Yukawa coupling constants,
$y_1$ and $y_2$.
The Lagrangian~\eqref{LY} may be enforced,
for instance,
through an $S_3$ permutation symmetry among the $\alpha$ indices
together with the six $\mathbbm{Z}_2$ symmetries~\cite{GL2006}
\bs \label{xvcfd}
\ba
\zz_2^{(\alpha,1)}: && D_{\alpha L} \to - D_{\alpha L}, \quad
\phi_\alpha \to - \phi_\alpha, \quad
\nu_{\alpha R} \to - \nu_{\alpha R},
\\
\zz_2^{(\alpha,2)}: && \phi_\alpha \to -\phi_\alpha, \quad \alpha_R \to -\alpha_R.
\ea
\es
Both
$\zz_2^{(\alpha,1)}$ and $\zz_2^{(\alpha,2)}$ act trivially
on fields with flavour $\beta \neq \alpha$.

If one imposes the $CP$ symmetry
\bs
\label{cp}
\ba
D_L \left( x_0, \vec x \right) &\to& i S \gamma_0 C
\bar D_L^T \left( x_0, -\vec x \right),
\label{bjwet} \\
\ell_R \left( x_0, \vec x \right) &\to& i S \gamma_0 C
\bar \ell_R^T \left( x_0, -\vec x \right)
\label{ivut}
\\
\nu_R \left( x_0, \vec x \right) &\to& i S \gamma_0 C
\bar \nu_R^T \left( x_0, -\vec x \right),
\label{vtyi} \\
\phi \left( x_0, \vec x \right)
&\to& S \phi^\ast \left( x_0, -\vec x \right),
\\
\phi_\nu \left( x_0, \vec x \right)
&\to& \phi_\nu^\ast \left( x_0, -\vec x \right),
\ea
\es
then one obtains real $y_1$ and $y_2$.
In transformation~\eqref{cp},
\be
D_L = \left( \begin{array}{c}
  D_{eL} \\ D_{\mu L} \\ D_{\tau L}
\end{array} \right),\
\ell_R = \left( \begin{array}{c}
  e_R \\ \mu_R \\ \tau_R
\end{array} \right),\
\nu_R = \left( \begin{array}{c}
  \nu_{eR} \\ \nu_{\mu R} \\ \nu_{\tau R}
\end{array} \right),\
\phi = \left( \begin{array}{c}
  \phi_e \\ \phi_\mu \\ \phi_\tau
\end{array} \right).
\ee
Clearly,
the transformation~\eqref{bjwet} includes, in particular,
the transformation~\eqref{bugiou}.
Let the VEVs be
\be
\left\langle 0 \left| \phi_l \right| 0 \right\rangle = \frac{v_l}{\sqrt{2}}
\left( \begin{array}{c} 0 \\ 1 \end{array} \right)
\ee
for $l = e,\mu,\tau,\nu$.
The charged-lepton mass matrix is diagonal:
\be
M_\ell = \frac{y_1}{\sqrt{2}}
\left( \begin{array}{ccc} v_e & 0 & 0 \\ 0 & v_\mu & 0 \\
  0 & 0 & v_\tau \end{array} \right).
\ee
The breaking of $CP$ is spontaneous,
via $v_\mu \neq v_\tau$,
corresponding to $m_\mu \neq m_\tau$.

In order to complete the model,
one introduces the bare Majorana mass terms
\be
\mathcal{L}_\mathrm{Maj} = \frac{1}{2}\, \nu_R^T C^{-1} M_R^* \nu_R + 
\mbox{H.c.}
\ee
and one assumes a (type~1) seesaw mechanism~\cite{seesaw}.
Clearly,
the neutrino Dirac mass matrix is
$M_D = \left( y_2^\ast v_\nu \left/ \sqrt{2} \right. \right) \bone$ and
\be
M_\nu = - M_D^T M_R^{-1} M_D = - \frac{{y_2^\ast}^2 v_\nu^2}{2}\, M_R^{-1}.
\ee
One further requires that in $\mathcal{L}_\mathrm{Maj}$
\begin{itemize}
\item the $\zz_2^{(\alpha,1)}$ are explicitly but \emph{softly}\/ broken
  through a non-diagonal $M_R$,
\item the $CP$ symmetry~(\ref{vtyi}) is \emph{conserved}.
\end{itemize}
Due to the $CP$ symmetry one has
\be
\label{mrs}
M_R = S M_R^\ast S = \left( \begin{array}{ccc} r & c^\ast & c \\
  c^\ast & c^\prime & r^\prime \\
  c & r^\prime & {c^\prime}^\ast
\end{array} \right),
\ee
where $r$ and $r^\prime$ are real while $c$ and $c^\prime$ are complex.
Equation~(\ref{mu-tau}) then holds,
because $M_\nu \propto M_R^{-1}$.
One thus has a model 
corresponding to the first approach of section~\ref{cobi}.
Note that the neutrino masses are not predicted in this model.

We now seek to transform this model into an \emph{equivalent}\/ one
that accords with the second approach of the previous section.
In equation~(\ref{LY}),
the Yukawa-coupling matrices $\Gamma_l$
responsible for the charged-lepton masses are
\be \label{gfyth}
\Gamma_e = \left( \begin{array}{ccc} 
y_1 & 0 & 0 \\ 0 & 0 & 0 \\ 0 & 0 & 0 
\end{array} \right),
\
\Gamma_\mu =  \left( \begin{array}{ccc} 
0 & 0 & 0 \\ 0 & y_1 & 0 \\ 0 & 0 & 0 
\end{array} \right),
\
\Gamma_\tau = \left( \begin{array}{ccc} 
0 & 0 & 0 \\ 0 & 0 & 0 \\ 0 & 0 & y_1 
\end{array} \right),
\
\Gamma_\nu = \left( \begin{array}{ccc} 
0 & 0 & 0 \\ 0 & 0 & 0 \\ 0 & 0 & 0 
\end{array} \right),
\ee
in the notation
\be
\label{gygut}
\mathcal{L}_Y = - \sum_{\alpha, \beta = e, \mu, \tau}\ \sum_{l = e, \mu, \tau, \nu}
\bar D_{\alpha L} \left[ \phi_l \left( \Gamma_l \right)_{\alpha \beta} \beta_R
  + \tilde \phi_l \left( \Delta_l \right)_{\alpha \beta} \nu_{\beta R} \right]
+ \mathrm{H.c.}
\ee
The other Yukawa-coupling matrices are
\be
\Delta_e = \Delta_\mu = \Delta_\tau = 0, \quad
\Delta_\nu = y_2 \bone.
\ee
Now we transform the fields
into a basis where the $CP$ symmetry has the standard form.
Comparing equations~(\ref{cp}) and~\eqref{buigfp},
one sees that
\be
S_D = S_\ell = S_\nu = S, \quad
S_\phi = \left( \begin{array}{cc} 
S & 0_{3 \times 1} \\ 0_{1 \times 3} & 1
\end{array} \right).
\ee
We want to use equation~\eqref{bufiy}
to obtain $S^\prime_D = S^\prime_\ell = S^\prime_\nu = \bone$
and $S^\prime_\phi = \mathrm{diag} \left( 1, 1, 1, 1 \right)$.
Equation~\eqref{bone} suggests
to choose\footnote{In equation~\eqref{nihjo} we might have utilized $U_\varrho$
  instead of $U_\omega$.
  Indeed, we are free to utilize \emph{any}\/ matrix $U$
  that satisfies $S U = U^\ast$ instead of $U_\omega$ in equation~\eqref{nihjo}.
  If we had utilized $U_\varrho$, we would have obtained the matrices
  $\Gamma^\prime_{\mu,\tau}$ in equation~\eqref{bjpec} below,
  with $y_3 \to y_1$.}
\be \label{nihjo}
W_D = W_\ell = W_\nu = U_\omega, \quad
W_\phi = \left( \begin{array}{cc} 
U_\omega & 0_{3 \times 1} \\ 0_{1 \times 3} & 1
\end{array} \right),
\ee
with $U_\omega$ of equation~(\ref{Uomega}).
One may now write the Yukawa-coupling matrices in the new basis
by using equations~\eqref{bugfvr}.
One obtains
\be \label{bvphoui}
\Gamma^\prime_e = \frac{y_1}{\sqrt{3}}\, \bone, \quad
\Gamma^\prime_\mu = \frac{y_1}{\sqrt{3}}\, E^T, \quad
\Gamma^\prime_\tau = \frac{y_1}{\sqrt{3}}\, E, \quad
\Delta^\prime_\nu = y_2 \bone,
\quad \mbox{where} \
E \equiv \left( \begin{array}{ccc}
0 & 1 & 0 \\ 0 & 0 & 1 \\ 1 & 0 & 0
\end{array} \right).
\ee
Note that $\Gamma^\prime_\nu = \Delta^\prime_e = \Delta^\prime_\mu
= \Delta^\prime_\tau = 0$. 
The Yukawa-coupling matrices are real
\emph{before}\/ and \emph{after}\/ the basis transformation
because of the $CP$ symmetry.
Obviously,
the Yukawa-coupling matrices look simpler
in the first approach---equation~\eqref{gfyth}---than
in the second approach---equation~\eqref{bvphoui}.
On the other hand,
$M^\prime_R = U_\omega M_R U_\omega$ is simply a real matrix in the second approach,
and therefore the neutrino Majorana mass terms
look simpler in the second approach than in the first one.

\section{Another simple model for cobimaximal mixing}
\label{simple model2}

In our second model the Yukawa Lagrangian is
\be
\mathcal{L}_Y =
- \bar D_{eL} \left( y_1 \phi_e e_R + y_2 \tilde \phi_\nu \nu_{eR} \right)
- \sum_{\alpha=\mu,\tau} \bar D_{\alpha L} 
\left( y_3 \phi_\alpha \alpha_R + 
y_4 \tilde\phi_\nu \nu_{\alpha R} \right) + \mbox{H.c.}
\ee
This Lagrangian may be enforced,
for instance,
through a $\mu$--$\tau$ permutation symmetry
together with the $\mathbbm{Z}_2$ symmetries~\eqref{xvcfd}.
We once again impose the $CP$ symmetry~\eqref{cp},
making $y_{1,2,3,4}$ real.
The charged-lepton mass matrix is
\be
M_\ell = \frac{1}{\sqrt{2}}
\left( \begin{array}{ccc} y_1 v_e & 0 & 0 \\ 0 & y_3 v_\mu & 0 \\
  0 & 0 & y_3 v_\tau \end{array} \right).
\ee
The breaking of $CP$ occurs via $v_\mu \neq v_\tau$.

We use once again the seesaw mechanism
with the matrix $M_R$ of equation~\eqref{mrs}.
We thus obtain another model for cobimaximal mixing,
very similar to the one of the previous section
but with a Yukawa Lagrangian with four coupling constants
instead of just two.

We next transform this model into a form
that accords with the second approach of section~\ref{cobi}.
The initial Yukawa-coupling matrices are
\be
\Gamma_e = \left( \begin{array}{ccc} 
y_1 & 0 & 0 \\ 0 & 0 & 0 \\ 0 & 0 & 0 
\end{array} \right),
\
\Gamma_\mu =  \left( \begin{array}{ccc} 
0 & 0 & 0 \\ 0 & y_3 & 0 \\ 0 & 0 & 0 
\end{array} \right),
\
\Gamma_\tau = \left( \begin{array}{ccc} 
  0 & 0 & 0 \\ 0 & 0 & 0 \\ 0 & 0 & y_3
\end{array} \right),
\
\Delta_\nu = \left( \begin{array}{ccc} 
y_2 & 0 & 0 \\ 0 & y_4 & 0 \\ 0 & 0 & y_4 
\end{array} \right),
\ee
and $\Gamma_\nu = \Delta_e = \Delta_\mu = \Delta_\tau = 0$.
We transform the fields
into a basis where the $CP$ symmetry has the standard form by using
\be \label{bohu}
W_D = W_\ell = W_\nu = U_\varrho, \quad
W_\phi = \left( \begin{array}{cc} 
U_\varrho & 0_{3 \times 1} \\ 0_{1 \times 3} & 1
\end{array} \right),
\ee
with $U_\varrho$ of equation~(\ref{Urho}).
Note that in equation~\eqref{bohu} we use the matrix
$U_\varrho$ instead of $U_\omega$ 
as in equation~\eqref{nihjo}.
This is just an arbitrary choice---we may use
any unitary matrix $U$ such that $S U = U^\ast$,
but using $U_\varrho$ in this case yields simpler results.
We write down the Yukawa-coupling matrices in the new basis:
$\Gamma_e$,
$\Gamma_\nu = 0$,
and the four $\Delta_l$ remain invariant while
\be
\label{bjpec}
\Gamma^\prime_\mu = \frac{y_3}{\sqrt{2}}
\left( \begin{array}{ccc} 0 & 0 & 0 \\ 0 & 1 & 0 \\ 0 & 0 & 1
\end{array} \right),
\quad
\Gamma^\prime_\tau = \frac{y_3}{\sqrt{2}}
\left( \begin{array}{ccc} 0 & 0 & 0 \\ 0 & 0 & -1 \\ 0 & 1 & 0
\end{array} \right).
\ee
The Yukawa-coupling matrices are still real,
but now this follows from the standard $CP$ symmetry.
The matrix $M^\prime_R = U_\varrho M_R U_\varrho$
is real too.

\section{Accommodation of the SM Higgs}
\label{accommodation}

We have two models,
both of them with four scalar doublets,
namely the $\phi_l$ with $l=e,\mu,\tau,\nu$.
We must use
those doublets
to give mass to the quarks. 
It suggests itself to use $\phi_\nu$ for this purpose.
Then $\left| v_\nu \right|$ should be large,
because the top-quark Yukawa coupling cannot be much larger than~1.
Consequently,
it is natural to assume
(for definiteness, in the model of section~\ref{simple model1})
\be \label{vevs-ineq}
\left| v_\alpha \right|^2 = \frac{2 m_\alpha^2}{y_1^2} \ll \left| v_\nu \right|^2
\approx \left( 246\, \mathrm{GeV} \right)^2.
\ee
(In this section,
as elsewhere in this paper,
the indices $\alpha$,
$\beta$,
and $\gamma$ vary in the range $\left\{ e, \mu, \tau \right\}$.)
Obviously,
the inequalities $\left| v_e \right|^2 \ll \left| v_\mu \right|^2
\ll \left| v_\tau \right|^2$
hold due to the strong hierarchy among the charged-lepton masses.

We know that a scalar field $h$ has been discovered at LHC.
We also know that the couplings of that scalar
to the \emph{heavy}\/ fermions and to the gauge bosons
are close to the couplings of the SM Higgs.
The couplings of $h$ to the light fermions
may be at variance with the ones in the SM.

We use the formalism in refs.~\cite{Grimus:1989pu,GL2002,Grimus:2007if}
for the neutral scalars.
The neutral components of the doublets $\phi_l$ are written
\be
\phi_l^0 = \frac{1}{\sqrt{2}} \left( v_l + \sum_{b=1}^8
\mathcal{V}_{lb} S_b^0 \right),
\ee
where the eight fields $S_b^0$ ($b = 1, 2, \ldots, 8$)
are neutral eigenstates of mass.
By definition,
$S_1^0$ is the neutral ``would-be'' Goldstone boson
and the $S_b^0$ for $b = 2, \ldots, 8$ are physical scalars.
The $4 \times 8$ complex matrix $\mathcal{V}$ has the property that
\be
\tilde{\mathcal{V}} = \left( \begin{array}{c} \mathrm{Re}\, \mathcal{V} \\
  \mathrm{Im}\, \mathcal{V} \end{array} \right)
\ee
is $8 \times 8$ orthogonal.
The first column of $\mathcal{V}$,
which corresponds to the Goldstone boson,
is given by $\mathcal{V}_{l1} = i v_l / v$,
where
\be
v = \sqrt{\left| v_e \right|^2 + \left| v_\mu \right|^2 + \left| v_\tau \right|^2
+ \left| v_\nu \right|^2}.
\ee
The interaction of the neutral scalars with a pair of SM gauge bosons
is given by the Lagrangian~\cite{Grimus:2007if}
\be \label{buighff}
\mathcal{L} = \cdots
+ g \left( m_W W_\mu^+ W^{\mu -} + \frac{m_Z Z_\mu Z^\mu}{2 c_w} \right)
\sum_{b=2}^8 S_b^0\ \mathrm{Im} \left( \mathcal{V}^\dagger \mathcal{V}
\right)_{b1}.
\ee
Equation~\eqref{buighff} indicates that we should have
$\mathrm{Im} \left( \mathcal{V}^\dagger \mathcal{V} \right)_{b1} = 1$
for a neutral scalar with index $b>1$ that has couplings
to pairs of gauge bosons identical to the ones of the SM Higgs.
Since $\mathcal{V}_{l1} = i v_l / v$,
$\mathrm{Im} \left( \mathcal{V}^\dagger \mathcal{V} \right)_{b1} = 1$
is equivalent to $\mathcal{V}_{lb} = v_l / v$.
Thus,
in order to exactly reproduce the SM Higgs,
the vector
\be \label{smallv}
\mathbf{v} = \frac{1}{v} \left( \begin{array}{c}
  v_e \\ v_\mu \\ v_\tau \\ v_\nu
\end{array} \right)
\ee
must be one of the columns of the matrix $\mathcal{V}$,
namely the one corresponding to a neutral-scalar mass eigenstate
with mass $m_h = 125$\,GeV.
We discuss here the possibility that $\mathbf{v}$
is an \emph{exact}\/ mass eigenstate.
One may eventually consider an approximate mass eigenstate,
taking into account the inequality~(\ref{vevs-ineq}).

The conditions that the scalar potential must satisfy
in order for the vector $\mathbf{v}$ 
to correspond to a neutral mass eigenstate with mass $m_h$ are given
by equations~\eqref{ab}.
It is easy to fulfil equation~(\ref{a}) by finetuning.
One simply has to assume 
for the matrix $\mu^2$ in the quadratic part of the scalar potential that 
\be \label{buih}
\mu^2 = - \frac{m_h^2}{2}\, \mathbf{v} \mathbf{v}^\dagger + \cdots,
\ee
where the dots indicate a part of the matrix $\mu^2$ operating
in the space orthogonal to $\mathbf{v}$.

We make the following simplifying assumptions
concerning the scalar potential $V = V_2 + V_4$:
\begin{itemize}
\item In the quartic part $V_4$ we assume invariance
  under any permutation of $e$,
  $\mu$,
  and $\tau$;
  in addition,
  we assume that all the doublets occur in pairs
  $\phi_l$, $\phi_l^\dagger$.
  Thus,
  \ba
  V_4 &=& \lambda\, \sum_\alpha \left( \phi_\alpha^\dagger \phi_\alpha \right)^2
  + \lambda_\nu \left( \phi_\nu^\dagger \phi_\nu \right)^2
  \no & &
  + \frac{1}{2}\, \sum_{\alpha \neq \beta} \left[
    \lambda' \left( \phi_\alpha^\dagger \phi_\alpha \right)
    \left( \phi_\beta^\dagger \phi_\beta \right) +
    \lambda'' \left( \phi_\alpha^\dagger \phi_\beta \right)
    \left( \phi_\beta^\dagger \phi_\alpha \right) \right]
  \no & &
  + \sum_\alpha \left[
    \lambda'_\nu \left( \phi_\nu^\dagger \phi_\nu \right)
    \left( \phi_\alpha^\dagger \phi_\alpha \right)
    + \lambda''_\nu \left( \phi_\nu^\dagger \phi_\alpha \right)
    \left( \phi_\alpha^\dagger \phi_\nu \right) \right].
  \label{v4}
  \ea
  This may be achieved by observing that the Yukawa Lagrangian
  in equation~\eqref{LY} is invariant under
  \be
  \label{duigho}
  \begin{array}{l}
  \left( \begin{array}{c}
    D_{eL} \\ D_{\mu L} \\ D_{\tau L}
  \end{array} \right) \to \left( \begin{array}{c}
    D_{eL} \\ \omega^2 D_{\mu L} \\ \omega D_{\tau L}
  \end{array} \right), \quad
  \left( \begin{array}{c}
    e_R \\ \mu_R \\ \tau_R
  \end{array} \right) \to \left( \begin{array}{c}
    e_R \\ \omega \mu_R \\ \omega^2 \tau_R
  \end{array} \right),
  \\*[8mm]
  \left( \begin{array}{c}
    \nu_{eR} \\ \nu_{\mu R} \\ \nu_{\tau R}
  \end{array} \right) \to \left( \begin{array}{c}
    \nu_{eR} \\ \omega^2 \nu_{\mu R} \\ \omega \nu_{\tau R}
  \end{array} \right), \quad
  \left( \begin{array}{c} \phi_e \\ \phi_\mu \\ \phi_\tau
  \end{array} \right) \to \left( \begin{array}{c}
    \phi_e \\ \omega \phi_\mu \\ \omega^2 \phi_\tau
    \end{array} \right),
  \end{array}
  \ee
  and then by requiring $V_4$ to be invariant under this transformation too.
\item We allow for a general quadratic part $V_2$.
  This breaks softly both symmetries~\eqref{xvcfd} and~\eqref{duigho}.
  The soft symmetry breaking also occurs in the matrix
  $M_R$ anyway,
  therefore we must assume its presence in $V_2$ too.
\end{itemize}
In order to use the notation of equation~\eqref{fye} we identify,
in the $V_4$ of equation~\eqref{v4},
\bs
\ba
\lambda_{\alpha\alpha\alpha\alpha} &=& \lambda,
\\
\lambda_{\nu\nu\nu\nu} &=& \lambda_\nu, 
\\
\lambda_{\alpha\alpha\beta\beta} = \lambda_{\beta\beta\alpha\alpha}
&=& \frac{\lambda'}{2} \quad (\alpha \neq \beta),
\\
\lambda_{\alpha\beta\beta\alpha} = \lambda_{\beta\alpha\alpha\beta}
&=& \frac{\lambda''}{2} \quad (\alpha \neq \beta), 
\\
\lambda_{\alpha\alpha\nu\nu} = \lambda_{\nu\nu\alpha\alpha}
&=& \frac{\lambda'_\nu}{2}, 
\\
\lambda_{\nu\alpha\alpha\nu} = \lambda_{\alpha\nu\nu\alpha}
&=& \frac{\lambda''_\nu}{2},
\ea
\es
whence we find
\bs \label{Lambda}
\ba
\Lambda_{\alpha\alpha} &=& 
\lambda \left| v_\alpha \right|^2
+ \frac{\lambda'}{2} \sum_{\gamma \neq \alpha} \left| v_\gamma \right|^2
+ \frac{\lambda'_\nu}{2} \left| v_\nu \right|^2, 
\\
\Lambda_{\nu\nu} &=& 
\lambda_\nu \left| v_\nu \right|^2
+ \frac{\lambda'_\nu}{2} \sum_{\gamma} \left| v_\gamma \right|^2, 
\\
\Lambda_{\alpha\beta} &=& \frac{\lambda''}{2}\, v_\beta^* v_\alpha
\quad (\alpha \neq \beta),
\\
\Lambda_{\nu \alpha}^\ast = \Lambda_{\alpha\nu}
&=& \frac{\lambda''_\nu}{2}\, v_\nu^* v_\alpha.
\ea
\es
Therefore,
\bs
\ba
\left( \Lambda \mathbf{v} \right)_\nu &=&
\left( \lambda_\nu \left| v_\nu \right|^2
+ \frac{\lambda'_\nu + \lambda''_\nu}{2} \sum_\beta \left| v_\beta \right|^2
\right) \frac{v_\nu}{v},
\\
\left( \Lambda \mathbf{v} \right)_\alpha &=&
\left( \lambda \left| v_\alpha \right|^2
+ \frac{\lambda' + \lambda''}{2} \sum_{\beta \neq \alpha} \left| v_\beta \right|^2
+ \frac{\lambda'_\nu + \lambda''_\nu}{2} \left| v_\nu \right|^2
\right) \frac{v_\alpha}{v}.
\ea
\es
Equation~(\ref{b}) then reads
\bs
\label{vmgkp}
\ba
\frac{m_h^2}{2} &=&
\lambda_\nu \left| v_\nu \right|^2
+ \frac{\lambda'_\nu + \lambda''_\nu}{2} \sum_\beta
\left| v_\beta \right|^2,
\label{e1} 
\\
\frac{m_h^2}{2} &=&
\lambda \left| v_\alpha \right|^2
+ \frac{\lambda' + \lambda''}{2} \sum_{\beta \neq \alpha} \left| v_\beta \right|^2
+ \frac{\lambda'_\nu + \lambda''_\nu}{2} \left| v_\nu \right|^2.
\label{vyep}
\ea
\es
Equation~\eqref{vyep} separately holds for $\alpha = e, \mu, \tau$.

The right-hand side of equation~\eqref{vyep} has,
in general,
a dependence on $\alpha$,
while its left-hand side is independent of $\alpha$.
Consistency is achieved by assuming
$\lambda = \left.\left( \lambda^\prime + \lambda^{\prime \prime} \right) \right/ 2$.

Equations~\eqref{vmgkp} may be satisfied by assuming
a custodial-type symmetry~\cite{custodial} in $V_4$.
Since we have four Higgs doublets, we may choose $U(4)$. 
The $U(4)$-symmetric quartic potential is then 
\be
\tilde V_4 = a \left(\, \sum_{l=e,\mu,\tau,\nu} \phi_l^\dagger \phi_l \right)^2 + 
b \sum_{l=e,\mu,\tau,\nu} \, \sum_{l^\prime=e,\mu,\tau,\nu}
\left( \phi_l^\dagger \phi_{l^\prime} \right) 
\left( \phi_{l^\prime}^\dagger \phi_l \right).
\ee
Comparison of $\tilde V_4$ with
$V_4$ of equation~\eqref{v4} yields
\be
\lambda = \lambda_\nu = a + b, \quad
\lambda' = \lambda'_\nu = 2 a, \quad
\lambda'' = \lambda''_\nu = 2 b.
\ee
Equations~(\ref{e1}) and~(\ref{vyep}) then merge into 
\be
\frac{m_h^2}{2} = \left( a + b \right) v^2.
\ee

\section{Conclusions}
\label{concl}

Neutrino oscillation data indicate that 
cobimaximal mixing may be a viable scenario,
at least as a first approximation,
for lepton mixing. In the literature there are two approaches to 
cobimaximal mixing.
They may be characterized by the underlying $CP$
symmetry of the mass Lagrangian of the light neutrinos.
The first approach features a generalized $CP$ symmetry that
includes a $\mu$--$\tau$ interchange,
in the basis where the charged-lepton mass matrix is diagonal.
In the second approach,
the $CP$ symmetry is the standard one,
but the charged-lepton mass matrix is non-diagonal and 
provides a factor $U_{\ell L}^\dagger$ in the mixing matrix 
$V = U_{\ell L}^\dagger U_\nu$ such that $S U_{\ell L}^\dagger = U_{\ell L}^T$,
with $S$ given by equation~\eqref{S}.
We have demonstrated that the two approaches yield the same consequences 
for lepton mixing,
on the one hand by using a simple mathematical theorem
which proves the equivalence of the mixing matrices in both
approaches,
and on the other hand by explicitly stating
the weak-basis transformation
that transforms the generalized $CP$ symmetry into the 
standard one.

Moreover,
we have displayed two renormalizable models for cobimaximal mixing 
which illustrate the relationship between the two approaches. 
In these seesaw models, the mixing angles $\theta_{12}$ and 
$\theta_{13}$ as well as the neutrino masses are undetermined. They use not 
only the above-mentioned $CP$ symmetry but also flavour symmetries which are 
softly broken in the Majorana mass terms of the right-handed neutrino singlets;
in this way,
cobimaximal mixing is obtained straightforwardly.
Finally, since each of our models has four Higgs doublets, 
the accommodation of the SM Higgs boson is non-trivial.
Using the general discussion of this issue for any number of Higgs doublets 
presented in appendix~\ref{VEVa},
we have formulated some necessary conditions for the 
existence of a neutral scalar
with SM-like couplings in the two models.

\newpage

\appendix

\section{Basis transformation}
\label{bt}
\setcounter{equation}{0}
\renewcommand{\theequation}{A.\arabic{equation}}

Let the Yukawa Lagrangian be as in equation~\eqref{gygut}.
We perform the basis transformation
\be \label{bughihjp}
D_L = W_D D^\prime_L, \quad \ell_R = W_\ell \ell^\prime_R, \quad
\nu_R = W_\nu \nu^\prime_R, \quad \phi = W_\phi \phi^\prime.
\ee
The transformed Yukawa-coupling matrices are
\be \label{bugfvr}
\Gamma^\prime_l = \sum_{l^\prime} \left( W_\phi \right)_{l^\prime l}
W_D^\dagger \Gamma_{l^\prime} W_\ell,
\quad
\Delta^\prime_l = \sum_{l^\prime} \left( W_\phi^\ast \right)_{l^\prime l}
W_D^\dagger \Delta_{l^\prime} W_\nu.
\ee
Let the $CP$ symmetry be
\be \label{buigfp}
D_L \to i S_D \gamma_0 C \bar D_L^T, \quad
\ell_R \to i S_\ell \gamma_0 C \bar \ell_R^T, \quad
\nu_R \to i S_\nu \gamma_0 C \bar \nu_R^T, \quad
\phi \to S_\phi \phi^\ast.
\ee
The $CP$-transformation matrices in the new basis are
\be \label{bufiy}
S'_D = W_D^\dagger S_D W_D^*, \quad
S'_\ell = W_\ell^\dagger S_\ell W_\ell^*, \quad
S'_\nu = W_\nu^\dagger S_\nu W_\nu^*, \quad
S'_\phi = W_\phi^\dagger S_\phi W_\phi^*.
\ee

\newpage

\section{VEV alignment for the SM Higgs}
\label{VEVa}
\setcounter{equation}{0}
\renewcommand{\theequation}{B.\arabic{equation}}

We use in this appendix the notation and results
of appendix~A of ref.~\cite{GL2002}.
Suppose there are $n_H$ Higgs doublets $\phi_i$
($i = 1, 2, \ldots, n_H$).
The scalar potential is
\be \label{fye}
V = \sum_{i,j=1}^{n_H} \mu^2_{ij} \left( \phi_i^\dagger \phi_j \right)
+ \sum_{i,j,k,l=1}^{n_H} \lambda_{ijkl} \left( \phi_i^\dagger \phi_j \right)
\left( \phi_k^\dagger \phi_l \right) \equiv V_2 + V_4,
\ee
where $V_2$ is the quadratic part of the potential
and $V_4$ is the quartic part.
From now on we employ the summation convention.
We define the matrices $\Lambda$,
$K$,
and $K^\prime$ through
\be
\label{bfsigp}
\Lambda_{ij} \equiv \lambda_{ijkl} v_k^\ast v_l, \quad
K_{ik} \equiv \lambda_{ijkl} v_j v_l, \quad
K^\prime_{il} \equiv \lambda_{ijkl} v_j v_k^\ast.
\ee
We also define the vector
\be
\label{v}
\mathbf{v} \equiv
\left( \sum_{i=1}^{n_H} \left| v_i \right|^2 \right)^{-1/2}
\left( \begin{array}{c} v_1 \\ v_2 \\ \vdots \\ v_{n_H}
\end{array} \right).
\ee
It is clear from these definitions that
\be
\label{fugtp}
\Lambda \mathbf{v} = K \mathbf{v}^\ast = K^\prime \mathbf{v}.
\ee
The expectation value of the potential in the vacuum state is
\be
\label{vigeo}
V_0 = \frac{1}{2}\, \mu^2_{ij} v_i^\ast v_j
+ \frac{1}{4}\, \lambda_{ijkl} v_i^\ast v_j v_k^\ast v_l.
\ee
Since the vacuum state is a stationary point of $V_0$,
\be
0 = \frac{\partial V_0}{\partial v_i^\ast} = \frac{1}{2} \left(
\mu^2_{ij} v_j + \lambda_{ijkl} v_j v_k^\ast v_l \right)
\ee
(note that $\lambda_{ijkl} = \lambda_{klij}$ by definition).
We thus have
\be\label{1}
\left( \mu^2 + \Lambda \right) \mathbf{v} = 0.
\ee

As explained after equation~\eqref{smallv},
we want $\mathbf{v}$ to be a column of the matrix $\mathcal{V}$
corresponding to a neutral scalar with mass $m_h$,
therefore it has to
fulfil
(see equation~(A18) of ref.~\cite{GL2002})
\be \label{2}
\left( \mu^2 + \Lambda + K^\prime \right) \mathbf{v} + K \mathbf{v}^\ast =
m_h^2 \mathbf{v}.
\ee
Because of equation~\eqref{fugtp},
equation~\eqref{2} may be rewritten
\be \label{3}
\left( \mu^2 + 3 \Lambda \right) \mathbf{v} = m_h^2 \mathbf{v}.
\ee
We then employ equation~\eqref{1} to obtain
\bs
\label{ab}
\ba
\mu^2 \mathbf{v} &=& - \frac{m_h^2}{2}\, \mathbf{v},
\label{a}
\\
\Lambda \mathbf{v}  &=& \frac{m_h^2}{2}\, \mathbf{v}.
\label{b}
\ea
\es

\end{document}